\documentclass[rmp,aps,preprint,groupedaddress]{revtex4}      
\begin{document}
\def\mathbolditalic#1{\textbf{\em #1}}
\title{Axiom  system   and  completeness   expression  for   quantum  mechanics}
\author{Carsten Held} 
\affiliation{Philosophisches Seminar, Universit\"{a}t Erfurt,
Postfach  90  02  21, D-99105  Erfurt,  Germany} 
\date{05.16.2007} 
\begin{abstract} 
\noindent
The  standard
axiomatization of quantum mechanics (QM) is not fully explicit about the role of
the time-parameter. Especially, the time reference within the probability
algorithm (the Born Rule, BR) is unclear. Using a plausible principle P1, about
the role of probability in a physical theory, and a second principle P2 affording a most natural
way to make BR precise, a logical conflict with the standard
expression for the completeness of QM can be derived. Rejecting P1 is implausible. Rejecting P2
leads to unphysical results and to a conflict with a generalization of P2, a
principle P3. It is thus made plausible that the standard expression of QM
completeness must be revised. An absolutely explicit form of the axioms is
provided, including a precise form of the projection postulate. An
appropriate expression for QM completeness, reflecting the restrictions of the
Gleason and Kochen-Specker theorems is proposed.
\end{abstract}
\maketitle
\tableofcontents

\section{Introduction}
\label{1}

Quantum mechanics (QM), the most elementary of quantum theories, can be shown to
be complete in a quite precise sense. 
\itshape
It is impossible to assign pre-existing values to suitable physical systems beyond the 
QM allowances, under two plausible constraints.
\normalfont
Take a physical system S such that its QM 
representation requires a Hilbert space $\mathcal{H}$ with dim($\mathcal{H}$)~$>$~2. (A one-particle spin-1 
system is an example.) It is impossible, in this case, to assign pre-existing values to all 
QM observables under the constraints that (i) algebraic relations among such value assignments 
mirror the algebraic relations among the (operators representing the) QM observables; 
(ii) the value assignments are non-contextual, i.e. every submaximal (degenerate) 
observable gets assigned a unique value, not one relativized to the different maximal 
(non-degenerate) observables of which it is a function. An 
\emph{indirect}
proof of this fact is immediately obtained from Gleason's Theorem [1]. 
The theorem entails that, when \{P\} is the set of all projection operators 
on $\mathcal{H}$, every mapping 
$\mu$: \{P\} $\rightarrow$ [0, 1] being interpretable as a probability function must 
be continuous, while a value assignment obeying (i) and (ii) must induce a mapping
$\mu^\prime$: \{P\} $\rightarrow$ [0, 1]
that is discontinuous [2]. A 
\itshape
direct
\normalfont
, i.e. constructive proof is the Kochen-Specker 
theorem [3, 4, 5, 6], presenting a finite set of operators for which an assignment obeying 
(i) and (ii) fails.\\

But why does the impossibility of assigning pre-existing values under these
constraints tell us anything about QM completeness? After all, the theory's
empirical output consists just in probabilities for measurement results and
their generalizations: expectation values. In an axiomatic formulation, QM is
formally incapable of directly making value assignments, so it cannot generate
anything conflicting with any value assignment to S. The
natural idea filling this logical gap is the insight that some probability
assignments entail value assignments, namely those that predict values with
certainty. E.g., a QM prediction to the effect that S, with probability 1, will
be found to have a property \emph{a}$_k$ at time t makes it plausible to conclude  that
S, at that time, has \emph{a}$_k$. Working, from now on, in the Schr\"{o}dinger picture,
writing states as density operators, and taking 
\bf \emph{A}
\normalfont
as a discrete observable on S
with values a$_1$, a$_2$, \ldots, we can express this idea as:\\

If \bf{P}\normalfont$_{a_k}$ (t), then \bf \emph{a}\normalfont $_k$ (t).\\

(Here, \bf{P}\normalfont$_{a_k}$ (t) = $|$ a(t) $>$ $<$ a(t) $|$, `\bf{P}\normalfont$_{a_k}$ (t)' 
abbreviates `S is in state \bf{P}\normalfont$_{a_k}$ (t)',
and `\bf \emph{a}\normalfont $_k$ (t)' abbreviates `S has \emph{a}$_k$ at t'.) 
Adding such a plausible rule to the QM formalism, we can extract value assignments, 
but nothing near a set of values big enough to conflict with either the discontinuous assignment 
used in the corollary of Gleason's Theorem or the assignment to some Kochen-Specker set of operators. 
This will be the case only if we \emph{limit} ourselves to the value assignments following from the QM state as follows:\\

\begin{tabbing}
  11111\=11111111111\=11111111111111111111111111111111111111111111\kill
\>\bf 	EE 	\>\bf\emph{a}\normalfont $_k$(t) if and only if \bf{P}\normalfont$_{a_k}$(t).\\
\end{tabbing}

This condition establishes a logical link between QM and the two theorems and thus makes precise in which sense 
they prove QM completeness. Indeed, the condition (often called the eigenstate-eigenvalue link, 
hence the label `\bf EE\normalfont ') embodies the classic definition of QM completeness [7, 8]. \bf EE \normalfont
substantiates the generally 
accepted and most familiar idea that a QM system in a superposition of \bf \emph{A}
\normalfont 
eigenstates does not have 
a value of \bf \emph{A}\normalfont. This idea plays a special role when interpreters try to say what goes on when S, being 
in  superposition of \bf \emph{A}\normalfont -eigenstates, meets an \bf \emph{A}\normalfont-measurement device. It is standardly claimed that S, 
during the measurement interaction, takes on one of the \bf \emph{A}
\normalfont 
values, e.g. \emph{a}$_k$ [9]. If S is found 
to have a value of \bf \emph{A}\normalfont, e.g. \emph{a}$_k$, at a certain time, then \bf EE \normalfont 
dictates that S's state is the 
pertaining eigenstate, e.g. \bf{P}\normalfont$_{a_k}$, at this time. A slightly less exact form of this consequence 
would be: If S is found to have value \emph{a}$_k$, of \bf \emph{A}\normalfont, then S's state immediately becomes the pertaining 
eigenstate e.g. \bf{P}\normalfont$_{a_k}$. This latter requirement is generally called the \emph{projection postulate} [10]. 
Projection, i.e. S's adopting an \bf \emph{A}\normalfont-eigenstate during \bf \emph{A}\normalfont-measurement, is generally thought to be an 
empirically confirmed fact and with good reason. We can measure copies of S for \bf \emph{A}\normalfont, filter out the 
non-\bf \emph{a}\normalfont $_k$ results, and then experimentally confirm the remaining state to be \bf{P}\normalfont$_{a_k}$, e.g. via quantum-state 
tomography [11, 12]. \bf EE \normalfont 
seems to have an exact version of the projection postulate as its consequence, 
thus seems to embody \emph{both} the completeness of QM and the necessity of projection upon finding a certain 
result. For future reference, I extract from \bf EE \normalfont 
the parts representing, respectively, the completeness 
condition (COMP) and the simplest candidate for a precise projection postulate (CPP):\\

\begin{tabbing}
  11111\=11111111111\=11111111111111111111111111111111111111111111\kill
\>\bf COMP 	\>\normalfont If S is not in state \bf{P}\normalfont$_{a_k}$(t), then not \bf \emph{a}\normalfont $_k$(t).\\
\\
\>\bf CPP	\>\normalfont If \bf \emph{a}\normalfont $_k$(t), then S is in state \bf{P}\normalfont$_{a_k}$(t).\\
\end{tabbing}

(Note that, by contraposition, both COMP and CPP express the same (backward or `only if'-) 
direction of \bf EE\normalfont.) The aim of the present paper is to show that COMP (and, consequently CPP) is in 
harmony with QM, in its standard axiomatization, and to provide more appropriate expressions for 
completeness and projection. More exactly, I will show COMP to be in conflict with QM as follows. 
I briefly review the standard axioms of QM and point out that two of them are not fully unambiguous concerning the 
role of the time parameter. I introduce three reasonable principles, P1-P3, where P3 is a generalized version of P2. 
The first principle P1 concerns the interpretation of probabilities in a physical theory, in general, 
while P2 and P3 refer specifically to the time parameter in QM and remove the ambiguity in the axioms. 
(All axioms and principles are introduced in Sec.~II.) However, using P1 and P2 to 
interpret QM probabilities, we can produce a contradiction from QM and COMP (Sec.~III). It will immediately be clear that  
P1 is not open to reasobale doubt and that sacrificing P2 leads to an implausible and unphysical consequence. 
Hence, the standard way to express the completeness of QM, i.e. COMP, must be revised. The latter 
is not an appropriate expression of the limitations generated by the Gleason and Kochen-Specker theorems. 
Principle P2 suggests a more precise version of the QM axioms, including a precise version of 
the projection postulate (Sec.~VI). Finally, a version of completeness will be proposed that both represents the 
limitations due to the two theorems and respects these axioms (Sec.~VII).\\

Some interpretations of QM reject the projection postulate and \bf EE\normalfont ; they are now collected under 
the title of \emph{modal interpretations} [13]. This group of interpretations has a weaker expression of 
completeness at hand:\\

\begin{tabbing}
  11111\=11111111111\=11111111111111111111111111111111111111111111\kill
\>\bf COMP* 	\>\normalfont If S is in a pure state W(t) $\neq$  \bf{P}\normalfont$_{a_k}$(t), then not \bf \emph{a}\normalfont $_k$(t).\\
\end{tabbing}

The rationale of COMP* is this: While a measurement may leave S in a mixture 
(obtained by partial tracing of the state of the S-cum-apparatus-supersystem) 
such that we can say that S has adopted one of \bf \emph{A}\normalfont's values without state projection, 
we still can express the idea that S, in a pure state W(t) at interaction onset, does not 
have any value of \bf \emph{A}\normalfont. This possibility of implementing a weaker form of completeness into a weaker 
version of QM must be considered, which is done in a digression comprising Sec.s IV. and V. I show that 
rejecting P2, despite its plausibility must go along with rejecting the projection postulate, thus the acceptance 
of a reduced version of QM (the standard axioms without projection postulate), like the one adopted in modal 
interpretations (Sec.~IV). Then I consider the third principle P3, a generalized version of P2, and show that 
COMP* and the reduced version of QM make P3 implausible (Sec.~V). By nature of the foundational and conceptual 
questions involved, the reasoning will consist of logical, not mathematical, argument throughout.

\section{Axioms and Principles}
\label{2}

Consider the following standard axiomatization of QM, using again the Schr\"{o}dinger picture and projection operators:\\

\begin{tabbing}
  11111\=11111111111\=11111111111111111111111111111111111111111111\kill
\>\bf A1 	\>\normalfont Any QM system S is associated with a unique Hilbert space $\mathcal{H}$ and its\\
\>		\>state is represented by a unique density operator W(t) on $\mathcal{H}$, a function\\
\>		\>of time.\\
\\
\>\bf A2 	\>\normalfont Any physical quantity \bf{A} \normalfont (called an observable) is represented by a\\
\>		\>self-adjoint operator \bf \emph{A} \normalfont on $\mathcal{H}$ and the possible values of \bf{A} \normalfont(possible\\ 
\>		\>properties of S) by the numbers in the spectrum of \bf \emph{A}\normalfont.\\
\\
\>\bf A3 	\>\normalfont S evolves in time according to W(t) $=$ U(t)W(t$_0$)U(t)$^{-1}$ where\\
\>		\>U(t)$=$exp$[$-i\bf\emph{H}\normalfont t$]$, a unitary operator, is a function of time and \bf \emph{H} \normalfont is an\\
\>		\> operator representing the total energy of S.\\
\\
\>\bf A4 	\>\normalfont If S is in state W(t) and \bf \emph{A} \normalfont is an observable on S, then the\\
\>		\>expectation value $<$\bf \emph{A}\normalfont$>$ is: $<$\bf \emph{A}\normalfont$>$ = Tr(W(t)\bf\emph{A}).\\
\\
\>\bf A5 	\>\normalfont If S is found to have value \emph{a}$_k$ as a result of an \bf \emph{A} \normalfont measurement, then\\ 
\>		\>S's state is \bf{P}\normalfont$_{a_k}$ immediately after this measurement.\\
\end{tabbing}

In view of the above discussion, I henceforth denote as QM the theory based on A1-A5, 
while a reduced version, based on A1-A4 only, will be called QM$^-$. Axiom A2 motivates 
an identification of physical observables and their mathematical representatives and I 
will not need to distinguish them. I will also, for simplicity, restrict myself to one 
discrete observable \bf \emph{A} \normalfont throughout. Finally, I will mostly restrict A4 to probabilities, i.e. 
expectation values of yes-no observables of type \bf P\normalfont $_{a_i}$. Let \emph{a}$_k$ always be some fixed value of 
variable \emph{a}$_i$. Let `p(\bf \emph{a}\normalfont $_k$)' mean the probability that S has \emph{a}$_k$. 
Then, since $<$\bf{P}\normalfont$_{a_k}>$ = p(\bf \emph{a}\normalfont $_k$), A4 
takes on a simpler, very familiar form, called the Born Rule (BR):\\

\begin{tabbing}
  11111\=11111111111\=11111111111111111111111111111111111111111111\kill
\>\bf BR 	\>\normalfont If S is in state W(t) and \bf \emph{A} \normalfont is an observable on S with eigenvalue \emph{a}$_k$,\\
\>		\>then the probability that S has \bf \emph{a}\normalfont $_k$ is: p(\bf \emph{a}\normalfont $_k$) = Tr(W(t)\bf{P}\normalfont$_{a_k}$).\\
\end{tabbing}

It should be emphasized that these axioms, though fairly standard, do not constitute a fully satisfactory 
axiomatization of QM since A4 and A5, in their present form, leave the role of the time parameter 
unspecified or vague. The defect in A4 carries over to BR, in whose equation only the right side, but not the left, 
carries a time-index. Two of the three principles, to be discussed presently, will have the sole purpose of forcing 
an unambiguous explication of the time-parameter on the left side of `p(\bf\emph{a}\normalfont $_k$) = Tr(W(t)\bf P\normalfont$_{a_k}$)' and it should be stressed 
that the interpretations produced from these principles and considered below exhaust all the reasonable options.\\

Here are three principles, the first concerning the role of probability in a physical theory in general, 
the second and third its role in QM. The first principle can be motivated by the idea that probability is 
quantified possibility. More precisely: If a physical theory assigns an event a non-zero probability, then, 
given the theory's truth, this event is possible. The weakest form of possibility is logical possibility. Thus, 
yet more precisely:\\

\begin{tabbing}
  11111\=11111111111\=11111111111111111111111111111111111111111111\kill
\>\bf P1 	\>\normalfont If, for a proposition F (describing an event) a theory T yields another\\
\>		\>proposition p(F) $>$ 0, then it is not the case that T, F $\vdash \perp$.\\
\end{tabbing}

(Here `T, F $\vdash \perp$' means that the set of sentences including F and all sentences of 
T allows to derive a contradiction in first-order logic.) P1 is beyond reasonable doubt, but it also follows 
from natural assumptions about probability shared by the main interpretations of that notion [14].\\

The second principle runs:\\

\begin{tabbing}
  11111\=11111111111\=11111111111111111111111111111111111111111111\kill
\>\bf P2 	\>\normalfont Any expression `\bf \emph{a}\normalfont $_k$' such that it names a QM event can be qualified as\\
\>		\>`\bf \emph{a}\normalfont $_k$(t)', where t is a time-parameter.\\
\end{tabbing}

P2 is motivated by the idea that a fundamental physical theory must explicitly concern spacetime events. 
A fundamental theory that builds probability spaces over sets of events must be able to explicitly treat 
these events as spacetime events. Hence, all events that are assigned probabilities in QM must explicitly be 
spacetime events, here: properties (like \emph{a}\normalfont $_k$) possessed at certain sharp times. Fully relativistic versions of 
QM explicitly treat spacetime events with a finite time-extension $\Delta$t. In the present, non-relativistic, 
formulation we have  $\Delta$t =  $\delta$t: QM events just consist in S having one or more properties at a sharp time t. 
It should be stressed that relativistic generalizations always contain the limiting case $\delta$t (see, e.g. [15]), hence 
an argument affecting QM in this respect will affect any relativistic generalization. Note, however, that P2 just says 
that those events denoted by statements of type `\bf \emph{a}\normalfont $_k$' within the QM probabilities are spacetime events such that 
the expressions can be explicated as `\bf \emph{a}\normalfont $_k$(t)'. The `\bf \emph{a}\normalfont $_k$' may not be appropriate 
expressions of QM events within BR 
and P2 may have no application.

The third principle, P3, generalizes P2. It says, in effect, that whatever the QM events are 
(and whatever expressions denote them) these events can be qualified as spacetime events explicitly. Thus:\\

\begin{tabbing}
  11111\=11111111111\=11111111111111111111111111111111111111111111\kill
\>\bf P3 	\>\normalfont For any expression `F' such that it QM yields an expression\\
\>		\>`p(F) = Tr(W(t)\bf{P}\normalfont$_{a_k}$)' there is a parameter t in the formalism qualifying\\
\>		\> `F' as `F(t)'.\\
\end{tabbing}

P3 is formulated so wide as to appear vague. But it has only two specifications. The first is to place the 
time-index `inside the probability', the second `outside the probability'. Consider the BR 
expression `p(\bf \emph{a}\normalfont $_k$) = Tr(W(t)\bf{P}\normalfont$_{a_k}$)' made precise as `p(\bf \emph{a}\normalfont $_k$(t))~=~Tr(W(t)\bf{P}\normalfont$_{a_k}$)'. This makes QM fulfill P2. 
But there is an alternative: Read the BR expression as `p(t)(\emph{a}$_k$)~=~Tr(W(t)\bf P\normalfont$_{a_k}$)' 
and interpret the latter in the following way: The probability is a disposition of S at time t to display value 
\emph{a}$_k$ (make `\bf \emph{a}\normalfont $_k$' true). This idea is discussed widely in the literature and is generally explicated by saying that t 
is the onset time of a measurement interaction on S and p(t)(emph{a}$_k$)' quantifies S's strength of disposition at t 
toward displaying \emph{a}\normalfont $_k$ at some later time [4]. However, while this notion essentially refers to the idea of 
probabilities as dispositions it does not need to refer to measurement. We should avoid the impression that 
anything in our principles or axioms makes essential reference to measurement - as this is in fact unnecessary. 
We can speak more generally of a region of space containing S and a time t such that `E(t)' names the disposition 
at t to display \emph{a}\normalfont $_k$ at some later time and call `E(t)' the triggering event. We can then 
say that an alternative to 
the preceding is to disambiguate `p(\bf \emph{a}\normalfont$_k$)' as `p(t)(\emph{a}$_k$)', which more explicitly 
reads `p(\bf \emph{a}\normalfont$_k$)~given~E(t)). We thus 
have an alternative disambiguation of BR that obeys our principle P3. (Note that the argument for 
only two possibilities 
is not strict. It would require heavy metalinguistic machinery to show 
that `p (\bf \emph{a}\normalfont $_k$(t))' and `p(t)(\bf \emph{a}\normalfont $_k$)' are the 
\emph{only} ways to specify the time-reference in `p(\bf \emph{a}\normalfont $_k$)'.)

\section{QM $+$ COMP contradict Principles P1 and P2}
\label{3}

P1 and P2 now generate the main argument against COMP. Using P2, we can make BR precise in a most natural way. 
It can now be rendered more exactly:\\

\begin{tabbing}
  11111\=11111111111\=11111111111111111111111111111111111111111111\kill
\>\bf BR$^\prime$ 	\>\normalfont If S is in state W(t) and \bf \emph{A} \normalfont is an observable on S with eigenvalue \emph{a}\normalfont $_k$,\\ 
\>			\>then the probability that S has \emph{a}$_k$ at t is: p(\bf \emph{a}\normalfont $_k$(t)) = Tr(W(t)\bf{P}\normalfont$_{a_k}$).\\
\end{tabbing}

Now suppose that S is in a state W(t$_1$) $\neq$ \bf P\normalfont $_{a_k}$ (t$_1$), for some value t$_1$ of t, such that from BR$^\prime$ 
it follows that 1 $>$ p(\bf\emph{a}\normalfont$_k$ (t$_1$)) $>$ 0. (Call this assumption N.) Assuming that a theory contains all 
its consequences, QM $+$ P2 will contain BR$^\prime$. Now, let QM $+$ P2, COMP, and N be integrated into one artificial 
theory, QM$^\prime$. Then, by simple sentential logic:\\

\begin{tabbing}
  11111\=111111111111111\=11111111\=1111111111111111111111111\=11111111111\kill

  \>N		\>(1)		\>S is in state W(t$_1$)		\>(N)\\
  
  \>N, BR$^\prime$ \>(2)	\>p(\bf \emph{a}\normalfont$_k$(t$_1$)) $>$ 0 \>(1), (BR$^\prime$)\\
  
  \>N		\>(3)		\>$\neg$ \bf{P}\normalfont$_{a_k}$(t$_1$) \>(N)\\
  
  \>N, COMP	\>(4)		\>$\neg$ \bf \emph{a}\normalfont$_k$ (t$_1$) \>(3), (COMP)\\
  
\end{tabbing}  
  
(As usual, the rightmost column indicates the assumptions on which the line in question directly depends and the leftmost 
column the ones on which the line ultimately depends.) By assumption, BR$^\prime$, COMP, N, are members of QM$^\prime$ which 
thus entails both line (2), i.e. that a certain proposition is assigned a positive probability, and line 
(4), i.e. that the negation of that proposition is true. Hence, QM$^\prime$ entails p(\bf\emph{a}\normalfont $_k$(t$_1$)) $>$ 0, 
but also: QM$^\prime$, \emph{a}$_k$(t$_1$) $\vdash$ $\perp$, in contradiction with P1. Thus given P1, QM$^\prime$ cannot be true [16].\\

The argument presupposes that QM$^\prime$, the artificial integration of QM, P2 and assumption N is a \emph{theory}. 
Is the integration of N an innocuous step? Of course, we can add suitable propositions to QM to create 
a theory that contradicts virtually any other proposition. But N is a trivially admissible state assignment 
that QM must be consistent with. So, its integration into QM$^\prime$ is innocuous indeed, but the one of P2 is not. 
BR$^\prime$, COMP, N are in conflict with P1, where BR$^\prime$ is BR, interpreted via P2. Given that P1 is immune to rejection, 
QM is in conflict with either P2 or COMP. 

\section{QM$^-$ as the Sole Alternative}
\label{4}

To reject P2 implies to give up on the most natural disambiguation of BR. The defender of COMP will just 
say that within the BR equation `p(\bf\emph{a}\normalfont $_k$)' cannot be read as `p(\bf\emph{a}\normalfont$_k$(t))', the impression of naturalness 
notwithstanding. But to reject P2 has consequences for the axioms. Recall that A5, like A4 and BR, is vague. 
Let's initially apply P2 to A5, yielding:\\

\begin{tabbing}
  11111\=11111111111\=11111111111111111111111111111111111111111111\kill
\>\bf A5$^\prime$ 	\>\normalfont If S is found to have value \emph{a}$_k$(t) as a result of an \bf \emph{A} \normalfont measurement, then\\
\>			\>S's state is \bf{P}\normalfont$_{a_k}$ immediately after this measurement.\\
\end{tabbing}

Note that A5$^\prime$ still substantially differs from CPP. Both, however, make precise the vague A5. 
Rejecting P2 would mean that A5 is not so made precise. It would mean, in effect, that expressions 
like `\bf\emph{a}\normalfont$_k$' are not explicated as `\bf\emph{a}\normalfont$_k$(t)' throughout QM. In this case, A5 automatically becomes vacuous. 
In the Schr\"{o}dinger picture, state evolution cannot start without a precise input state. It is the intention 
of A5, to generate such an input -- for starting post-measurement state evolution, e.g. when a measurement 
is a preparation. CPP generates a precise state from the precise `\bf\emph{a}\normalfont$_k$(t)' and A5$^\prime$ at least can be imagined 
to do so, when the phrase `immediately after' is made precise. If A5 is understood as containing an 
expression `\bf\emph{a}\normalfont$_k$' that must not carry a time-index, it has no such quality. It is a vacuous statement, 
not only without any empirical content, but also a formally ineffective addition to the rest of QM. So, 
everyone seeking to escape the argument of Sec.~III by rejecting P2 will have to reject the projection 
postulate in any substantial form.\\

One might object that applying P2 to BR is one thing and applying it to A5 another. 
But if we reject P2 for the expression `p(\bf\emph{a}\normalfont$_k$)' (refuse to read it as `p(\bf\emph{a}\normalfont$_k$(t))') we say that 
these probabilities do not mean probabilities for `\bf\emph{a}\normalfont$_k$(t)', nor that they are are they tested by 
observations of type `\bf\emph{a}\normalfont$_k$(t)'. It is inconsistent then to allow such an observation nevertheless 
and put it in the antecedent of A5.\\

All in all, rejecting P2 must go hand in hand with rejecting A5 and the interpretations taking this 
route are the modal interpretations. Here we must not consider this group of interpretations, in general, 
but a queer and artificial variant built on negating P2. Note that $\neg$ P2 immediately transforms QM into an 
unphysical theory. Checking the axioms of QM, we note that the theory (reasonably enough) contains a \emph{unique} 
time parameter. (The same goes for QM$^-$.) If the theory supplies a time-index for `\bf\emph{a}\normalfont$_k$' in `p(\bf\emph{a}\normalfont$_k$)', to obey P2, 
it must be this one. If, vice versa, we claim that `p(\bf\emph{a}\normalfont$_k$)' does not inherit the time-index directly from the 
state, i.e. from the right side of `p(\bf\emph{a}\normalfont$_k$) = Tr(W(t)\bf{P}\normalfont$_{a_k}$)', we automatically rule that it does not get any 
time-reference, at all. QM $+$ $\neg$ P2 does no longer furnish the measurement results, for which it provides probabilities, 
with exact time-indices. Perhaps we could come up with an additional theory of QM measurement fixing the problem, 
but any such theory would have to heavily revise the axioms, equipping the formalism with a second time-parameter.

\section{QM$^-$ $+$ COMP* make P3 implausible}
\label{5}

Notice that nothing in the previous argument hinges on whether S's state W(t) $\neq$ \bf{P}\normalfont$_{a_k}$(t) is a pure state 
or a mixture. So, if we can construct an argument similar to the one of Sec.~III, but referring to COMP* 
instead of COMP, we can disallow the combination QM$^-$ and COMP*. Such an argument can indeed be given using 
P3, but it lacks the rigor of the above one.\\

Since P2 alone forces the interpretation of BR as BR$^\prime$, the argument of Sec.~III can equally well be applied 
to COMP*. The defender of COMP* will have to reject P2 and revise BR$^\prime$. Given the assumption, made plausible 
above, that there is but one alternative way to specify BR, we will now rewrite it as:\\

\begin{tabbing}
  11111\=11111111111\=11111111111111111111111111111111111111111111\kill
\>\bf BR$^{\prime\prime}$ 	\>\normalfont If S is in state W(t) and \bf \emph{A} \normalfont is an observable on S with eigenvalue \emph{a}$_k$,\\
\>				\>then the probability that S has \emph{a}$_k$ given E(t) is:\\
\>				\>p(\bf \emph{a}\normalfont$_k$ given E(t)) = Tr(W(t)\bf{P}\normalfont$_{a_k}$).\\
\end{tabbing}

Assuming the triggering event E(t) to be the onset of an \bf \emph{A}\normalfont
-measurement interaction, 
we recover the idea, found in classical textbooks [17], that QM probabilities essentially are conditional upon 
measurement, and the idea that these probabilities are dispositions, possessed by S (or the whole of S and the 
apparatus) at time t, for S possessing \emph{a}$_k$ at some later time. However, as has been pointed out (at the end of 
Sec.~IV), this later time cannot be referred to in QM because the theory, as axiomatized here, does not have 
the formal resources to refer to two times. (Similarly, again, for QM$^-$.)\\

So, in the expression `\bf\emph{a}\normalfont$_k$ given E(t)' the `\bf\emph{a}\normalfont$_k$', referring to S and the time at which eventually 
it has \emph{a}$_k$, cannot bear a time-index. We have, thus, consciously violated P2, but not necessarily P3, 
since `\bf\emph{a}\normalfont$_k$ given E(t)' does contain some time reference, after all. As has been emphasized, however, the discussion 
at this point takes on an unphysical and academic character.\\

Probability expressions of the form `p(B given A) = z' (where z $\in$ [0, 1]) have been thoroughly 
investigated in the context of QM [18] and three possible analyses have been 
found: `p(B $|$ A) = z', `A $\rightarrow$ p(B) = z', and `p(A $\rightarrow$  B) = z', where `$\rightarrow$' is 
a conditional connective awaiting further semantic analysis. It should be added that philosophers 
and logicians have mounted substantial evidence that, in general, p(B $|$ A) $\neq$ p(A $\rightarrow$ B), for standard 
explications of `$\rightarrow$' [19]. So, these two forms of explicating `p(B given A) = z' are indeed logically 
different and we have (at least) three interpretations for the expression. In the present context, 
we have the special condition that `B' in `p(B given A) = z' must not bear a time-index, i.e. in the 
relevant BR$^{\prime\prime}$ expression `p(\bf\emph{a}\normalfont$_k$ given E(t)) = z' `\bf\emph{a}\normalfont$_k$' must not be time-indexed -- to escape the contradiction of Sec.~III.\\

We consider the three analyses of `p (\bf\emph{a}\normalfont$_k$ given E(t)) = z', in turn. It is easy to see 
that `p(\bf\emph{a}\normalfont$_k$ $|$ E(t)) = z' is not a live option. The standard (Kolmogorov) definition of 
conditional probability is inapplicable, since this would require `p(\bf\emph{a}\normalfont$_k$ $\land$ E(t))' and `p(E(t))' to be 
well-defined, which they are not. Defining them appropriately would mean to import them into QM from 
elsewhere -- something which is clearly inadmissible in a theory dubbed fundamental and, in addition, 
breaks the axiomatic closure of the theory. Alternatively, conditional probabilities can be defined as 
primitive two-place functions from pairs of events into the unit interval [20], but the axioms ruling the 
interpretation of these functions as probabilities require expressions like `p(E(t) $|$ \bf\emph{a}\normalfont$_k$)' to be well-defined. 
Again, no version of BR can supply such probabilities and importing them from elsewhere is out of the question.\\

Consider second `E(t) $\rightarrow$  p(\bf\emph{a}\normalfont$_k$) = z'. This variant contracts two problems. `z' is a placeholder 
for `Tr(W(t)\bf P\normalfont $_{a_k}$)', in BR$^{\prime\prime}$. Hence, we have the conditional 
`E(t)~$\rightarrow$~p(\bf\emph{a}\normalfont$_k$)~=~Tr(W(t)\bf P\normalfont$_{a_k}$)' 
containing, as its consequent, an equation 
`p(\bf\emph{a}\normalfont$_k$)~=~Tr(W(t)\bf P\normalfont$_{a_k}$)'. By assumption, this equation 
is no longer vague, but defined to lack a time index on the left and carry one on the right. For a 
mathematical function depending on some parameter, this is an inconsistent requirement. Moreover, 
exporting the time-reference from the set of events that get assigned probabilities via QM violates 
our principle P3.\\

Consider third p (E(t) $\rightarrow$ \bf\emph{a}\normalfont$_k$) = z. This possibility respects P3. 
But the unphysical assumption that its 
consequent `\bf\emph{a}\normalfont$_k$' must not bear a time-index makes it impossible to 
distinguish a case where `\bf\emph{a}\normalfont$_k$' is true 
at some unspecified time directly after E(t) from a case where `\bf\emph{a}\normalfont$_k$' is true at a 
much later time. This allows 
constructions of obviously false cases. Suppose that S is a one-particle 
spin-$\textstyle \frac{1}{2}$ system in W(t$_1$) = \bf P\normalfont$_{a_m}$(t$_1$), 
where \emph{a}$_m$ $\neq$ \emph{a}$_k$ is another eigenvalue of \bf \emph{A}\normalfont. Suppose that E(t$_1$) is the onset of a measurement interaction 
consisting in a series of measurements \bf \emph{A} \normalfont -- \bf \emph{B} \normalfont -- \bf \emph{A} \normalfont
(where [\bf \emph{A}\normalfont, \bf \emph{B}\normalfont] $\neq$ 0). Suppose that, despite the initial 
state \bf P\normalfont$_{a_m}$(t$_1$), the second \bf \emph{A}\normalfont -measurement yields result `\bf\emph{a}\normalfont$_k$'. 
Then `E(t$_1$)~$\rightarrow$~\bf\emph{a}\normalfont$_k$' is true and 
yet p(E(t$_1$) $\rightarrow$  \bf\emph{a}\normalfont$_k$) = Tr \bf P\normalfont$_{a_m}$(t$_1$)\bf P\normalfont$_{a_k}$ = 0. Of course, we will understand the physics of the experiment and 
say that the probability of `\bf\emph{a}\normalfont$_k$' being true directly after E(t) is zero and raises during the course of the 
whole experiment, but without a time reference we lack the possibility to distinguish different instances 
of `\bf\emph{a}\normalfont$_k$'. The point is not that we cannot come up with an intelligible distinction of instances 
of `\bf\emph{a}\normalfont$_k$', 
but rather that we cannot do so \emph{within} the present (mutilated) version of QM, where BR is interpreted as 
BR$^{\prime\prime}$ and , in turn, is read as delivering expressions of type 
`p(E(t) $\rightarrow$ \bf\emph{a}\normalfont$_k$) = Tr(W(t)\bf P\normalfont$_{a_k}$)'.\\

This argument for a violation of P3 is non-rigorous because it is built on two unproven meta-assumptions: (i) that 
the two disambiguations sketched in Sec.II and used at the beginning of this section are the only possible 
ones; (ii) that the three proposed analyses of `p(\bf\emph{a}\normalfont$_k$ given E(t)) = z' exhaust the possibilities. The 
argument for QM$^-$ violating P2 is non-rigorous, too, but is based upon a fairly trivial metalinguistic 
observation about the axioms: There is a unique time-index in A1--A4.

\section{Revised axioms for QM}
\label{6}

As a consequence of the preceding discussion, we cannot express the completeness of QM by COMP or COMP*. 
Moreover CPP, our simplest candidate for a precise version of the projection postulate A5, cannot be used 
to make it precise. But amendments to the axioms, guaranteeing harmony with P1-P3, are easily made. A set of axioms 
for QM respecting P1--P3 will consist of A1--A2 above plus A3*, A4*, and A5*, specified as follows:\\

\begin{tabbing}
  11111\=11111111111\=11111111111111111111111111111111111111111111\kill
\>\bf A3* 	\>\normalfont S evolves in time according to W(t) = U(t)W(t$_P$)U(t)$^{-1}$ where\\
\>		\>U(t) = exp[-i\bf \emph{H}\normalfont t], a unitary operator, is a function of time and \bf \emph{H} \normalfont is\\
\>		\>an operator representing the total energy of S, where $t_P$, some value of,\\
\>		\>t is called the preparation time, and W(t$_P$) the prepared state.\\
\\
\>\bf A4* 	\>\normalfont If S is in state W(t) $\neq$ W(t$_P$) and \bf \emph{A} \normalfont is an observable on S, then the\\
\>		\>expectation value $<$\bf \emph{A}\normalfont$>$(t) = $\int$ a(t)p(a(t))d$\omega$ is: $<$\bf \emph{A}\normalfont$>$(t) = Tr(W(t)\bf \emph{A}\normalfont).\\
\\
\>\bf A5*	\>\normalfont If S has value \emph{a}$_k$ (t$_1$) of \bf \emph{A}\normalfont, then t$_1$ = t$_P$ and S's state is the prepared\\
\>		\>state \bf{P}\normalfont$_{a_k}$(t$_P$).\\
\end{tabbing}

Some remarks on these axioms are in order. In A4*, the integral version of the expectation value serves to clearly 
specify the meaning of `$<$\bf \emph{A}\normalfont$>$ (t)'. It explicates that the events, the weights of which go into the QM 
expectation value, are time-indexed events of type `S has \emph{a}$_k$'. It can indeed be argued (and has been done elsewhere [21]) that 
requiring the QM expectations to be expected values, as they are usually defined in statistics, forces this 
form for them. A4* now allows deriving a final version of BR, BR*, via the familiar 
identification p(\bf\emph{a}\normalfont$_k$(t)) = $<$\bf P\normalfont$_{a_k}>$(t):\\

\begin{tabbing}
  11111\=11111111111\=11111111111111111111111111111111111111111111\kill
\>\bf BR* 	\>\normalfont If S is in state W(t) $\neq$ W(t$_P$) and \bf \emph{A} \normalfont is an observable on S with\\
\>		\>eigenvalue \emph{a}$_k$, then the probability that S has \emph{a}$_k$ at t is:\\
\>		\>p(\bf \emph{a}\normalfont$_k$(t)) = Tr(W(t)\bf{P}\normalfont$_{a_k}$).\\
\end{tabbing}

BR* is BR$^\prime$, with the restriction that W(t$_P$) is not an admissible input. More explicitly, 
BR* and A5* in conjunction rule that if `\bf\emph{a}\normalfont$_k$(t$_1$)' is 
true then no calculation of a number p(\bf\emph{a}\normalfont$_k$(t$_1$)) 
is allowed. This is not an implausible restriction. It is reasonable indeed to assume that the factual 
observation of an event at a certain time makes it meaningless to calculate any prediction for that event 
at that time. Note also that the only axiom making direct reference to the result of a factual observation, 
i.e. to a property ascription to S, is A5*. Note finally the three crucial virtues of this axiom system: 
(1) It is absolutely explicit concerning the time parameter; (2) it does not need to use the notion of 
measurement in any sense; (3) it allows us consistently to describe measurements as preparations because 
our findings upon measurement can be used, via A5*, as an input for A3*.

\section{An Expression of Completeness}
\label{7}

The completeness of QM is embodied in the theorems mentioned above, among others: 
the corollary from Gleason's Theorem and versions of the Kochen-Specker Theorem. 
We have seen that COMP is not an admissible way to express the impossibility results incorporated 
in these theorems. Our principle P1 embodies the most reasonable idea that probability is quantified 
possibility and P2--P3 represent plausible ways to render precise the imprecise A1--A5. Given these principles, 
COMP cannot be an expression of the impossibilty results, hence of the sense in which QM can be proved to be 
complete. But what \emph{is} an appropriate expression?\\

To repeat the first observation of this paper: 
\itshape
It is impossible to assign pre-existing values to suitable physical systems beyond the QM allowances, 
under two plausible constraints.
\normalfont 
We will now see that QM, made precise in the sense of A1--A5*, does yield probabilities for pre-existing values. 
Hence, it cannot be the idea of assigning pre-existing values as such, but the one of doing so under conditions 
(i) and (ii) which we should interpret as disproved by the completeness theorems. One or both of conditions (i) 
and (ii) for the assignment of pre-existing values must be rejected or modified.\\

It is easy to see that we have produced a general argument for the existence of pre-existing values. 
Consider, once more, S being in a state W(t$_1$) $\neq$ \bf P\normalfont$_{a_k}$(t$_1$) such that p(\bf\emph{a}\normalfont$_k$)
gets a value other than 
1 or 0, where t$_1$ is the onset time of an \bf \emph{A}\normalfont-measurement interaction. By BR*, `p(\bf\emph{a}\normalfont$_k$)' is explicated 
as `p(\bf\emph{a}\normalfont$_k$(t$_1$))', the probability that S has \emph{a}$_k$ at t$_1$, the onset time. So W(t$_1$), 
by our new axioms, 
collects probabilities for values possessed at the time of measurement onset, t$_1$. This is nothing but 
the assumption of pre-existing values. The rationale for BR* can be followed back into our principles. 
If `p(\bf\emph{a}\normalfont$_k$)' does not inherit the index t$_1$ it cannot bear any time-index, at 
all -- in contradiction with 
P2 and in obvious contrast with reasonable requirements for a fundamental probabilistic theory of 
spacetime events. If we sacrifice P2 nevertheless and take the remaining option for explicating a 
time-reference in `p(\bf\emph{a}\normalfont$_k$)', i.e. `p(\bf\emph{a}\normalfont$_k$ given E(t$_1$))', then 
no established construal of the conditional 
can both be coherent and respect P3. Respecting both P2 and P3, we end with BR*. Finally, if `(\bf\emph{a}\normalfont$_k$(t$_1$))' 
receives a positive probability, as it does in our case, it must be logically possible to assume it to be 
true. This is an instance of P1 and says that it must be logically possible to assume S having a 
value \emph{a}$_k$ of \bf \emph{A} \normalfont at t$_1$.\\

As a consequence, it cannot be true that QM is complete in the sense that the QM state W(t$_1$) provides 
all properties S has at t$_1$. Looking only at the axioms (here BR*), W(t$_1$) does nothing but collect 
probabilities for S's values at t$_1$. It is plausible to supplement the axioms with the rule that predictions 
with certainty entail value ascriptions (i.e. adopting the forward direction of \bf EE\normalfont : 
If \bf P\normalfont$_{a_k}$(t$_1$), then \bf\emph{a}\normalfont$_k$(t$_1$)), 
but it is implausible to bar all other ascriptions. 
Let \bf \emph{A} \normalfont and \bf \emph{B} \normalfont be discrete, with 
values \emph{a}$_1$, \emph{a}$_2$,$\ldots$, \emph{b}$_1$, \emph{b}$_2$,$\ldots$, 
and non-degenerate with [\bf \emph{A}\normalfont, \bf \emph{B}\normalfont] $\neq$ 0. 
Let S be in W(t$_1$) = \bf P\normalfont$_{b_j}$($t_1$). 
Then, by the rule just adopted, `\emph{b}$_j$(t$_1$)' is 
true and exactly one of `\emph{a}$_1$(t$_1$)', `\emph{a}$_2$(t$_1$)',$\ldots$ is true. Consider now a set of 
observables $\{$\bf P\normalfont$\}_{AB}$, that contains the 
projectors \bf P\normalfont$_{b_1}$, \bf P\normalfont$_{b_2}$,$\ldots$ \bf P\normalfont$_{a_1}$,
\bf P\normalfont$_{a_2}$,$\ldots$ and forms a Kochen-Specker set (i.e. a set such 
that a Kochen-Specker contradiction 
can be derived). What cannot be true, according to the Kochen-Specker Theorem, is that value assignments 
to all members 
of $\{$\bf P\normalfont$\}_{AB}$ do both of these two: (i) mirror the algebraic relations of the 
members of $\{$\bf P\normalfont$\}_{AB}$; 
(ii) are non-contextual, i.e.
are unique for every member of $\{$\bf P\normalfont$\}_{AB}$. There are, then, observables \bf \emph{A} \normalfont 
and \bf \emph{B} \normalfont such that all of the above assumptions 
are true, especially `\emph{b}$_j$(t$_1$)' is true and exactly one of `\emph{a}$_1$(t$_1$)', `\emph{a}$_2$(t$_1$)',$\ldots$ 
is true, and yet it \emph{cannot} be the 
case that of the \bf P\normalfont$_{b_1}$, \bf P\normalfont$_{b_2}$,$\ldots$ exactly one receives value 1, the others 0, 
and simultaneously, i.e. noncontextually, 
one of the \bf P\normalfont$_{a_1}$, \bf P\normalfont$_{a_2}$,$\ldots$ receives value 1, the others 0. In general, we arrive 
at the following completeness expression for QM:
\itshape
It is possible to assign pre-existing values to suitable physical systems beyond the QM allowances, 
but assignments cannot be such that values of submaximal (degenerate) observables mirror the 
algebraic relations among these observables noncontextually
\normalfont
[22, 23].\\

It is an open question what contextual value assignments would look like. As indicated, the 
context-dependence of pre-assigned values must be one of pre-existing values rather than one 
depending on measurement influences on S. The prospects for this type of contextuality 
(sometimes called `ontological contextuality') have been researched in the past [24, 25], 
but without much resonance. The present argument clearly shows that this possibility merits 
renewed attention. 

\begin{acknowledgements}

I am indebted to audiences at the Spring 2007 conference of the Deutsche 
Physikalische Gesellschaft at the Universit\"{a}t Heidelberg and at the 15$^{th}$
UK and
European Meeting on Foundations of Physics 2007 at the University of Leeds for
stimulating discussions.
\end{acknowledgements}


\begin{thebibliography}{99}
\bibitem{6} While the original argument [3] requires 117 projection operators, the simplest argument 
now can be given with only 18 projectors [4]. For a proof that this is the smallest possible set, see [5].
\bibitem{9} The classic proponent of this idea is, again, von Neumann. See [8], 206, 211.
\bibitem{14}Assume (P1 (a)) that contradictions have probability zero; (P1 (b)) the conditional 
probability formula: p(A $\land$ B) = p(A $|$ B) p(B); (P1 (c)) that the probability space for 
the probabilities delivered by T can be expanded so that p(T) is well-defined; and (P1 (d)) 
that p(T)~$>$~0. The non-trivial assumption is P1 (c). However, it can be made plausible for 
all major conceptions of probability. Consider probability being defined as a subjective 
degree of belief. (For a meaningful integration of this conception of probability into QM 
see [11].) Then it is rational to define p(T $|$ A) for a theory in order to be able to express 
that T's prediction A, if it comes out true, raises your degree of belief in it: p(T $|$ A) $>$ p(T). 
However, if p(T $|$ A) is well-defined, then p(A) and p(T) are well-defined on the same space. 
Consider, alternatively, probability being defined via conditional probabilities understood as 
ratios of proportions of logically possible worlds. p(T) then can be defined as p(T $|$ L) where 
L is a logical triviality and p(A $|$ T) is defined, on the same space as p(T), as the ratio of 
the proportion of logically possible T-worlds where A is true to the proportion of logically 
possible T-worlds. Consider, finally, probability being defined as the limiting relative frequency 
of possible outcomes in a hypothetical infinite sequence of trials of an experiment. Let a trial 
of an `experiment concerning T' be an explicit statement of T with possible `outcomes' True (T~=~1) 
and Not-true (T = 0). Then `T = 1' is an outcome as the event reported by A. Thus, p(T) can be 
defined as p(T = 1) on a superspace of the probability space where p(A) lives. Given P1 (a-d), 
the argument for P1 is very simple. Assume, by P1 (c) and P1 (d), that p(T)~$>$~0. Assume also that 
p(A $|$ T) $>$ 0. Then, by P1 (b), also p(A $\land$ T) $>$ 0, whence, by P1 (a), A $\land$ T is not a contradiction.
\bibitem{16}Alternatively, given P1 (a-c) from note [14], QM$^\prime$ cannot have a positive 
probability of being true.
\bibitem{23}The necessity to use sets containing submaximal (degenerate) observables to 
derive a Kochen-Specker contradiction is well-known since the work reported in [21].
\end{thebibliography}
\end{document}